%% file: main.tex
\documentclass[%
  aps,
  pra,
  reprint,
  superscriptaddress,
]{revtex4-2}

\usepackage[utf8]{inputenc}
\usepackage[english]{babel}

\usepackage{ifthen}
\usepackage[x11names]{xcolor}

\usepackage{amsmath}
\usepackage{bm}
\usepackage{mathtools}
\usepackage{siunitx}
\usepackage{xfrac}
\usepackage{calc}

\usepackage{graphicx}
\usepackage{tikz}

\usepackage[bookmarksopen]{hyperref}
\hypersetup{%
  colorlinks,
  linkcolor=SteelBlue4,
  citecolor=SteelBlue4,
  urlcolor=SteelBlue4,
}

\DeclareSIUnit\sccm{sccm}
\DeclareSIUnit\ppm{ppm}
\DeclareSIUnit\ppb{ppb}
\DeclareSIUnit\bar{bar}
\DeclareSIUnit\mbar{mbar}

\newcommand*{\pif}{University of Stuttgart, 5th Institute of Physics, Pfaffenwaldring 57, 70569 Stuttgart, Germany}
\newcommand*{\igm}{University of Stuttgart, Institute of Smart Sensors, Pfaffenwaldring 47, 70569 Stuttgart, Germany}
\newcommand*{\iis}{University of Stuttgart, Institute for Large Area Microelectronics, Allmandring 3b, 70569 Stuttgart, Germany}
\newcommand*{\density}{\ensuremath{N}}

\input{preamble/colors.tex}
\input{preamble/tikz.tex}

\input{preamble/macros.tex}

\newcommand*{\angfreq}[2]{\ensuremath{2\pi\times \SI{#1}{#2}}}

\begin{document}
\title{Collisional shift and broadening of Rydberg states in nitric oxide at room temperature}

\author{Fabian Munkes}
\affiliation{\pif}
\author{Alexander Trachtmann}
\affiliation{\pif}
\author{Patrick Kaspar}
\affiliation{\pif}
\author{Florian Anschütz}
\affiliation{\pif}
\author{Philipp Hengel}
\affiliation{\iis}
\author{Yannick Schellander}
\affiliation{\igm}
\author{Patrick Schalberger}
\affiliation{\igm}
\author{Norbert Fruehauf}
\affiliation{\igm}
\author{Jens Anders}
\affiliation{\iis}
\author{Robert Löw}
\affiliation{\pif}
\author{Tilman Pfau}
\affiliation{\pif}
\author{Harald Kübler}
\email{h.kuebler@physik.uni-stuttgart.de}
\affiliation{\pif}
%
\begin{abstract}
  We report on the collisional shift and line broadening of Rydberg states in nitric oxide (\no) with increasing density of a background gas at room temperature.
  As a background gas we either use \no~itself or nitrogen (\nitrogen).
  The precision spectroscopy is achieved by a sub-Doppler three-photon excitation scheme with a subsequent readout of the Rydberg states realized by the amplification of a current generated by free charges due to collisions.
  The shift shows a dependence on the rotational quantum state of the ionic core and no dependence on the principle quantum number of the orbiting Rydberg electron.
  The experiment was performed in the context of developing a trace--gas sensor for breath gas analysis in a medical application.
\end{abstract}
\maketitle

\input{content/introduction.tex}
\input{content/methods.tex}
\input{content/results.tex}
\input{content/conclusion.tex}

\bibliography{content/references.bib}

\end{document}

%% file: preamble/colors.tex
\definecolor{cmap0}{rgb}{0.001,0.000,0.013}
\definecolor{cmap1}{rgb}{0.019,0.015,0.088}
\definecolor{cmap2}{rgb}{0.066,0.038,0.186}
\definecolor{cmap3}{rgb}{0.129,0.047,0.290}
\definecolor{cmap4}{rgb}{0.197,0.038,0.367}
\definecolor{cmap5}{rgb}{0.271,0.040,0.411}
\definecolor{cmap6}{rgb}{0.341,0.062,0.429}
\definecolor{cmap7}{rgb}{0.403,0.085,0.433}
\definecolor{cmap8}{rgb}{0.472,0.110,0.428}
\definecolor{cmap9}{rgb}{0.540,0.134,0.415}
\definecolor{cmap10}{rgb}{0.603,0.157,0.395}
\definecolor{cmap11}{rgb}{0.670,0.184,0.366}
\definecolor{cmap12}{rgb}{0.735,0.215,0.330}
\definecolor{cmap13}{rgb}{0.791,0.250,0.291}
\definecolor{cmap14}{rgb}{0.846,0.297,0.244}
\definecolor{cmap15}{rgb}{0.894,0.353,0.193}
\definecolor{cmap16}{rgb}{0.929,0.411,0.145}
\definecolor{cmap17}{rgb}{0.959,0.482,0.089}
\definecolor{cmap18}{rgb}{0.978,0.557,0.034}
\definecolor{cmap19}{rgb}{0.986,0.630,0.030}
\definecolor{cmap20}{rgb}{0.986,0.713,0.103}
\definecolor{cmap21}{rgb}{0.974,0.797,0.206}
\definecolor{cmap22}{rgb}{0.956,0.874,0.323}
\definecolor{cmap23}{rgb}{0.947,0.949,0.491}
\definecolor{cmap24}{rgb}{0.988,0.998,0.644}

%% file: preamble/tikz.tex
\usetikzlibrary{calc}
\usetikzlibrary{3d}
\usetikzlibrary{arrows.meta}
\usetikzlibrary{bending}
\pgfdeclarelayer{bg2}
\pgfdeclarelayer{bg}
\pgfdeclarelayer{fg}
\pgfsetlayers{bg2,bg,main,fg}

%% file: preamble/macros.tex
\newboolean{ShowComments}\setboolean{ShowComments}{false}  
\provideboolean{ShowComments}
\newcommand{\ShowMyCommnt}[1]{%
  \ifShowComments%
    #1%
  \fi%
}
\newcommand{\NewComment}[2]{%
  \expandafter\newcommand\csname#1Comment\endcsname[1]{\ShowMyCommnt{\textcolor{#2}{##1}}}%
}
\newcommand{\FigRef}[1]{figure~\ref{#1}}

\newcommand{\EqnRef}[1]{equation~\ref{#1}}
\newcommand{\TblRef}[1]{table~\ref{#1}}

\NewComment{Harald}{violet}
\NewComment{Fabian}{cmap8}
\NewComment{Robert}{cmap15}

\newcommand*{\gs}{\ensuremath{\text{X}\,{}^2\Pi_{3/2}}}
\newcommand*{\gsgeneric}{\ensuremath{\text{X}\,{}^2\Pi}}

\newcommand*{\as}{\ensuremath{\text{A}\,{}^2\Sigma^{+}}}
\newcommand*{\hs}{\ensuremath{\text{H}\,{}^2\Sigma^{+}}}

\newcommand*{\ryd}{\ensuremath{n(N^{+})}}

\newcommand*{\tranUV}{\ensuremath{\as \leftarrow \gs}}
\newcommand*{\tranGreen}{\ensuremath{\hs \leftarrow \as}}
\newcommand*{\tranRed}{\ensuremath{\ryd \leftarrow \hs}}

\newcommand*{\nitrogen}{N\textsubscript{2}}
\newcommand*{\no}{NO}
\newcommand*{\cw}{cw}
\newcommand*{\tia}{TIA}
\newcommand*{\pcb}{PCB}
\newcommand*{\mfc}{MFC}
\newcommand*{\uv}{UV}
\newcommand*{\tisa}{Ti:sa}
\newcommand*{\ppln}{PPLN}
\newcommand*{\snr}{SNR}
\newcommand*{\aom}{AOM}
\newcommand*{\fwhm}{FWHM}


%% file: content/introduction.tex
\section{Introduction}
By the end of the 1980s it was known, that \no~plays an important role in the mammalian system \cite{Arnold1977,Furchgott1980,Ignarro1987}.
This lead to the Nobel prize in Physiology or Medicine being awarded to Murad, Furchgott and Ignarro in 1998 for their discovery \cite{NobelPrize}, which ultimately paved the way to a broad field of research \cite{Ignarro2018}, involving results on different forms of cancer \cite{Haklar2001,KordeChoudhari2012,XU2002,Khan2020} and the role of \no~in immunological responses such as inflammation \cite{Thomas2008}.
That \no~is also part of the exhaled breath, was shown in 1991 by Gustafsson et al.~\cite{Gustafsson1991}, and subsequent research revealed a change in the \no~concentration when diseases like asthma, atopy and others are present \cite{ATS2005a}.
The \no~concentration in the exhaled breath is in the low ppb--regime, and a guideline \cite{ATS2005a} suggests, that a sensor for breath--gas analysis must be calibrated with samples in the range of \SI{10}{\ppb} to \SI{100}{\ppb}.
Such a requirement is challenging when considering, that low gas volumes are preferred.
For example, \cite{ATS2005a} states, that a patient has to exhale for \SI{10}{\s} at a constant flow to gain about \SI{300}{\milli\l} of air volume for sensors readily available.
In the context of medical research such volumes are expected to be way lower.

We demonstrated a proof--of--concept experiment for a sensor based on the Rydberg excitation of \no~in \cite{Schmidt2018}, capable of detecting \no~concentrations less than $\SI{10}{\ppm}$ limited by preparation, and yet operable at ambient pressure.
The extrapolated sensitivity already reached the \SI{10}{\ppb} range.
In that experiment, only pulsed laser systems were used.
In the experiment presented in this work only continuous--wave (\cw) laser systems are used for the excitation of \no.
This ensures selective detection due to the linewidth of \cw~systems.
Our goal in this work is to learn about consequences for the sensor application.
As such we investigate collisional shifts and broadening of the Rydberg line with increasing density of the background gas.
This allows us to compare our results with previous results on the pressure broadening and shift in alkalis \cite{Fermi1934,Fuechtbauer1934,WeberNiemax1982}.
In a wider context, the consequential improvement in sensitivity may enable us to detect Rydberg bimolecules of \no~in the future, which have been predicted theoretically \cite{GonzlezFrez2021}.

%% file: content/methods.tex
\section{Methods}
\begin{figure}[b]
  \centering
  \input{content/setupFigure.tex}
  \caption{%
    Sketch of the setup and measurement principle used throughout this work.
    Gas mixtures of a certain concentration of \no~diluted in \nitrogen, or pure \no~enter the measurement cell on the left.
    Excitation of \no~to a Rydberg state is achieved by a \cw~three--laser excitation scheme shown on the upper left.
    Subsequent collisions ionize these molecules.
    Charges are collected via electrodes, and the current is amplified and converted to a voltage by employing a transimpedance amplifier (TIA).
    A lock-in amplifier referenced on the modulated second transition improves the signal--to--noise ratio (SNR).
    The cell's cuboid has a width of \SI{35}{\mm}, a depth of \SI{13.5}{\mm}, and a height of \SI{8.4}{\mm}.
    Pictures are shown in \FigRef{fig:picsCell}.
    \label{fig:setup}%
  }
\end{figure}
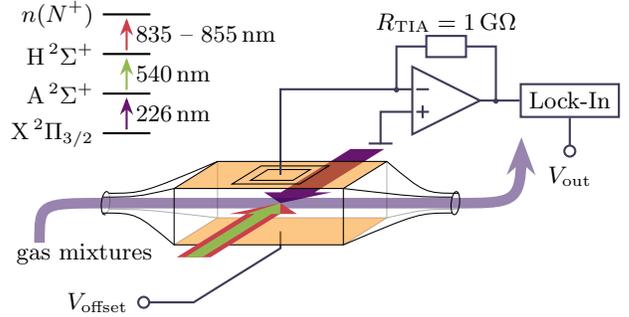

\paragraph*{Setup}
A sketch of the main components of the experimental setup is shown in \FigRef{fig:setup}.
All measurements are performed at room temperature, $T\approx\SI{293}{\kelvin}$.
The detection of Rydberg states in \no~is realized by the electronic detection of free charges.
As such the central part of our setup is a custom designed glass cell with built--in readout electronics.

\begin{figure}[b]
  \centering
  \includegraphics[width=\linewidth]{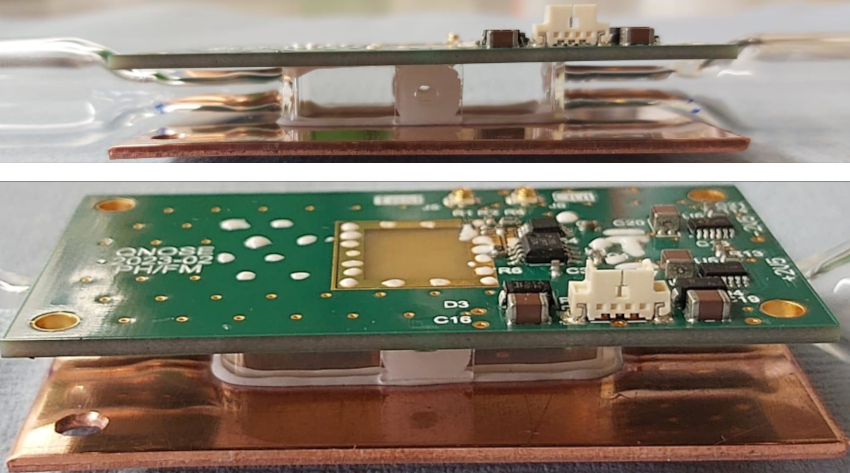}
  \caption{%
    Pictures of the readout cell.
    In the center of the upper photo we can see the quartz windows, which are glued to the glass frame, such that the UV--light may pass through.
    On the left and right are the attachments to the MFCs and pumps.
    The lower photo shows the PCB with readout electronics on top, as well as a copper plate on the bottom, which realizes the counterelectrode.
    \label{fig:picsCell}
  }
\end{figure}

The cell's glass frame has a copper plate glued to the bottom, whereas the top holds a printed-circuit board (\pcb) with an electrode pointing towards the cell's interior and amplification electronics on the other side.
Pictures are shown in \FigRef{fig:picsCell}.
Gas may flow through the cell as flange connectors are to the left and right of the frame.
Applying a (possibly small) potential to bottom and top electrodes collects free charges within the cell, and the needed amplification and conversion to a voltage of the current is achieved by the use of a transimpedance amplifier (\tia) with an overall feedback resistance of \SI{1}{\giga\ohm}.
The electrodes have an area of about \SI{472}{\mm\squared}.
While a constant and homogeneous field between the electrodes is desirable, a large area of the upper electrode connected to the \tia's input results in a large antenna collecting noise from surrounding sources.
Thus, a trade-off has to be made between the two.
Our solution is to divide the electrode up into sections, which we all keep at the same potential.
This is indicated in \FigRef{fig:setup} on the top electrode.
In our setup, the outermost part is a solder-resist free ground plane, then an actively driven guard ring follows, and the innermost area centered above the excitation volume is connected to the \tia.
This area has dimensions $\SI{13.5}{\mm} \times \SI{10}{\mm}$, where the longer side is parallel to the laser beams.

Pressure stability is achieved by using mass-flow controllers (\mfc) between the gas bottles and the experimental cell.
Two \mfc s are used for the experiment, one allows to regulate the flow of \no~in the range of \SI{0.1}{\sccm} to \SI{5}{\sccm}, and the other regulates the (optional) flow of \nitrogen~in a range of \SI{1.6}{\sccm} to \SI{100}{\sccm}.
If only pure \no~is used, the other \mfc~is closed off by valves.
The pressure is constantly monitored at both ends of the cell by pressure gauges.

The \no~molecules are excited to a Rydberg state by a three-photon excitation scheme solely by continuous-wave (\cw) laser systems shown on the top left in \FigRef{fig:setup}.
We only operate on transitions, where the vibrational quantum $v$ is $v=0$.
While the ultraviolet (\uv) beam enters the cell on the rear, the two other lasers are counter-propagating entering from the front.
Subsequent collisions of the excited \no~molecules with other particles yield free charges, which are electronically detected.
The \uv~system, a frequency-quadrupled titanium-sapphire laser (\tisa), was already introduced in our previous work \cite{Kaspar2022}, and is used to drive the first transition, $\tranUV$.
The $1/e^2$ beam waist $w$ is about \SI{940}{\micro\m}, and the mean intensity $I = P/\pi w^2$ is on average \SI{1.8}{\mW\per\mm\squared}.
The \cw~light for the second transition $\tranGreen$ is generated by a \SI{1080}{\nm} diode laser, which is fiber--amplified and finally frequency-doubled via a single-pass periodically-poled lithium niobate crystal (\ppln).
The beam has a waist of about \SI{640}{\micro\m} and an intensity of on average \SI{480}{\mW\per\mm\squared}.
The light for the Rydberg transition $\tranRed$ is another \tisa~with a beam waist of \SI{1220}{\micro\m} used at an intensity ranging from \SI{47}{\mW\per\mm\squared} to \SI{66}{\mW\per\mm\squared}.
Note that all beam waist measurements are the average value of values gathered \SI{10}{\cm} in front and after the cell.
The intensities are measured directly before entering the cell.

Our setup consists of a separate stabilization setup based on a reference laser at \SI{780}{\nm} locked to an ultra-low expansion cavity and then used to stabilize the length of transfer cavities.
This allows to lock the fundamental beams of our excitation lasers to their respective transfer cavities.
The Rydberg laser is not locked but scanned during the experiment, and its locked transfer cavity serves as a relative frequency reference.
The whole stabilization setup is based on several Red Pitaya STEMlab 125-14A and a self-written lock software built on top of PyRPL \cite{NeuhausPYRPL2017}.
A thorough walkthrough is given in \cite{laserLockTut}.
Additionally, all fundamental beams are sent to a wavemeter (HighFinesse WS-6) for monitoring.

An improvement in the signal--to--noise ratio (\snr) is gained by using a lock-in amplifier referenced by the modulation frequency of the second transition.
Modulation of the second transition itself is achieved by an acousto-optic modulator (\aom), which is driven by the amplified signal of a Red Pitaya STEMlab 125-14A.
\paragraph*{Experimental procedure}
After evacuating the cell by using a backing pump and a turbo pump to about \SI{1e-5}{\mbar} a constant flow of gas and thus a constant pressure is ensured by the \mfc s.
Two measurement types are performed, either using pure \no~or with mixtures of \no~and \nitrogen.
As soon as a stable pressure is achieved, the lasers for \tranUV~and \tranGreen~are locked to their respective branches.
For the \uv~transition this is the $\text{P}_{12} (6.5)$ branch at roughly $\SI{226.982}{\nm}$, and for the intermediate transition the $R_{11}(5.5)$ at about $\SI{540.494}{\nm}$ is used.
The wavelengths given are the wavemeter's readout, which has an inaccuracy of \SI{600}{\mega\hertz}.
The initial value of the \uv~transition was at first simulated with PGOPHER \cite{Western2017} by using constants from \cite{danielak1997}.
The initial wavelength of the branch $R_{11}(5.5)$ of the second transition was taken from \cite{Ogi2000}.

Finding the two lower and subsequently locked transitions is achieved by scanning and electronically reading out the signal as well.
Here, we use higher pressures, about \SI{1}{\mbar}, and higher fields, about \SI{10}{\V\per\cm}.
As soon as both transitions are locked the pressure is set a little lower to about \SI{60}{\micro\bar} and the applied field reduced to about \SI{-0.8}{\V\per\cm} to ease finding the Rydberg signal, since this avoids significant broadening and splitting.
The Rydberg signal is obtained by scanning the \tisa~operating on $\tranRed$.
Throughout the whole measurement we scan the Rydberg laser and trigger the scope on its ramp.
Finally, the pressure is set via the \mfc~to the desired start pressure.
Due to the positions of the gauges, the pressure inside the cell cannot be known exactly.
However, a bypass of the cell allows setting inbound and outbound pressure close to each other.
Thus, we assume the pressure inside the cell to be the mean of both measured values.

For our measurements we are able to choose the principal quantum number $n$ as well as the rotational quantum number of the ionic core $N^{+}$ of Rydberg states in \no.
A state is denoted by $n(N^+)$, and we omit specifying the ionic part $\text{X}^{+}\,{}^1\Sigma^{+}$, as it remains the same.
When a state is selected and all starting conditions are met the experimental data is acquired automatically.
Broken down to steps the procedure is:
\begin{enumerate}
  \item Set the applied field to $E_{\text{field}} \approx \SI{-11.9}{\V\per\cm}$, which results in a clear distinct splitting of the states due to the Stark effect.
  \item Adjust the measured frequency range such that the high-$l$-manifolds of the Stark effect are clearly visible.
        Here we take special care to make the dominant center peak of the manifold visible, which is the one targeted for the evaluation.
  \item Wait for \SI{45}{\s} to allow the flow (or pressure, respectively) to settle.
  \item During the scan the lock--in amplifiers output, the transmission signal of the transfer cavity of the Rydberg laser and, for checking, the trigger is recorded.
  \item When the scan is complete the flow is changed to the next value.
  \item Steps 3 to 5 are repeated.
\end{enumerate}
The measurement series has to be aborted if the reference cavity of the Rydberg laser goes out of lock, since we do not have an absolute frequency reference at hand.
Everything is acquired in `single-shot', i.e.~there is no averaging involved.
Additionally, we made sure to never reach the electronic limits, i.e.~keeping the scan velocity of the Rydberg laser such that it never outperforms the rise time of the electronic circuit and adjusting the lock--in amplifier's time constant accordingly.
In a single shot, we scanned over \SI{13}{\giga\hertz} in \SI{19.9}{\s}, which goes along with a lock-in time constant of \SI{30}{\milli\second} and a filter order of 3.
The rise time of the electronic circuit is below \SI{200}{\micro\s}, and as such negligible.

%% file: content/setupFigure.tex
\begin{tikzpicture}[%
    z={(-3.85mm,-2.4mm)}, 
  ]%
  \newlength\sizex\sizex=\linewidth
  \newlength\sizey\sizey=.666\linewidth
  \pgfmathsetmacro{\cellx}{1.5}
  \pgfmathsetmacro{\celly}{.5}
  \pgfmathsetmacro{\cellz}{1}
  \colorlet{colCellFrame}{cmap0}
  \colorlet{colLaserRed}{cmap13}
  \colorlet{colLaserGreen}{DarkOliveGreen3!90!black}
  \colorlet{colLaserUV}{cmap7}
  \colorlet{colElectrode}{cmap18}
  \colorlet{colGuard}{black}
  \colorlet{colWire}{cmap2!80!white}
  \colorlet{colGas}{cmap4}
  %
  \coordinate (origin) at (0,0,0);
  %
  %
  \begin{scope}[scale=1.5]
    \begin{scope}[draw=colCellFrame,line join=round]%
      \coordinate (ftl) at (-\cellx/2,\celly/2,\cellz/2); 
      \coordinate (fbl) at (-\cellx/2,-\celly/2,\cellz/2); 
      \coordinate (ftr) at (\cellx/2,\celly/2,\cellz/2); 
      \coordinate (fbr) at (\cellx/2,-\celly/2,\cellz/2); 
      \coordinate (rtl) at (-\cellx/2,\celly/2,-\cellz/2); 
      \coordinate (rbl) at (-\cellx/2,-\celly/2,-\cellz/2); 
      \coordinate (rtr) at (\cellx/2,\celly/2,-\cellz/2); 
      \coordinate (rbr) at (\cellx/2,-\celly/2,-\cellz/2); 
      \coordinate (ftleA) at (-\cellx/5,\celly/2,\cellz/3); 
      \coordinate (ftreA) at (\cellx/5,\celly/2,\cellz/3); 
      \coordinate (rtleA) at (-\cellx/5,\celly/2,-\cellz/3); 
      \coordinate (rtreA) at (\cellx/5,\celly/2,-\cellz/3); 
      \coordinate (ftleB) at (-\cellx/8,\celly/2,\cellz/5); 
      \coordinate (ftreB) at (\cellx/8,\celly/2,\cellz/5); 
      \coordinate (rtleB) at (-\cellx/8,\celly/2,-\cellz/5); 
      \coordinate (rtreB) at (\cellx/8,\celly/2,-\cellz/5); 
      \coordinate (fc) at ($(ftl)!.5!(fbr)$); front center
      \coordinate (cc) at ($(rtl)!.5!(rbr)$); front center
      \draw%
        (-\cellx/2,-\celly/2,\cellz/2) -- ++ (0,\celly,0)%
        --++(\cellx,0,0) -- ++(0,-\celly,0) --cycle;
      \draw%
        (-\cellx/2,\celly/2,\cellz/2) -- ++ (0,0,-\cellz)%
        -- ++(\cellx,0,0) --++(0,0,\cellz);

      \pgfmathsetmacro{\tuberadius}{.1}
      \begin{scope}[canvas is zy plane at x=\cellx/2+.5]
        \path (45:.1) coordinate (a);
        \path (135:.1) coordinate (b);
        \path (225:.1) coordinate (e);
        \path (315:.1) coordinate (c);
      \end{scope}

      \path (a) ++ (.3,0,0) coordinate (a2);
      \path (b) ++ (.3,0,0) coordinate (b2);
      \path (c) ++ (.3,0,0) coordinate (c2);
      \path (e) ++ (.3,0,0) coordinate (e2);

      \draw (ftr) .. controls (a) .. (a2);
      \draw (rtr) .. controls (b) .. (b2);
      \draw (fbr) .. controls (c) .. (c2);
      \begin{scope}[canvas is zy plane at x=\cellx/2+.8]
        \draw (0,0) circle (.1);
        \coordinate (gasend) at (0,0);
      \end{scope}
      %
      \begin{scope}[canvas is zy plane at x=-\cellx/2-.5]
        \path (45:.1) coordinate (a);
        \path (135:.1) coordinate (b);
        \path (225:.1) coordinate (d);
        \path (315:.1) coordinate (c);
      \end{scope}

      \path (a) ++ (-.3,0,0) coordinate (a2);
      \path (b) ++ (-.3,0,0) coordinate (b2);
      \path (c) ++ (-.3,0,0) coordinate (c2);
      \path (d) ++ (-.3,0,0) coordinate (d2);

      \draw (ftl) .. controls (a) .. (a2);
      \draw (rtl) .. controls (b) .. (b2);
      \draw (fbl) .. controls (c) .. (c2);
      \begin{scope}[canvas is zy plane at x=-\cellx/2-.8]
        \draw (0,0) ++ (135:.1)  arc (135:-90:.1);
        \coordinate (gasstart) at (0,0);
      \end{scope}
    \end{scope}
    \begin{pgfonlayer}{bg2}
      \begin{scope}[opacity=.3, transparency group]
        \draw (fbl) -- (rbl) -- (rbr) -- (fbr);
        \draw (rtl) -- (rbl);
        \draw (rtr) -- (rbr);
        \draw (rbl) .. controls (d) .. (d2);
        \begin{scope}[canvas is zy plane at x=-\cellx/2-.8]
          \draw (0,0) ++ (135:.1)  arc (135:270:.1);
        \end{scope}
        \begin{scope}
          \draw (rbr) .. controls (e) .. (e2);
        \end{scope}
      \end{scope}
    \end{pgfonlayer}
    %
    \begin{scope}
      \coordinate (cout) at (0,0,2);
      \coordinate (tbase) at ($(cout)!.8!(origin)$);
      \path (tbase)%
        +(.15,0,0) coordinate (tbaseright)
        +(.075,0,0) coordinate (tbaseright2)
        +(-.3,0,0) coordinate (tbaseleft)
        +(-.15,0,0) coordinate (tbaseleft2);
      \fill[fill=colLaserRed] (cout) ++(-.15,0,0) -- ++(.30,0,0) -- (tbaseright) -- ++(.15,0,0)
        -- (origin) -- (tbaseleft) -- ++ (.15,0,0) -- cycle;
      \fill[fill=colLaserGreen] (cout) ++(-.075,0,0) -- ++(.15,0,0) --
        (tbaseright2) -- ++(.075,0,0)
        -- (origin) -- (tbaseleft2) -- ++ (.075,0,0) -- cycle;
      \begin{pgfonlayer}{bg}
        \coordinate (cout) at (0,0,-2);
        \coordinate (tbase) at ($(cout)!.8!(origin)$);
        \path (tbase)%
          +(.15,0,0) coordinate (tbaseright)
          +(-.3,0,0) coordinate (tbaseleft);
        \fill[fill=colLaserUV] (cout) ++(-.15,0,0) -- ++(.30,0,0) -- (tbaseright) -- ++(.15,0,0)
          -- (origin) -- (tbaseleft) -- ++ (.15,0,0) -- cycle;
      \end{pgfonlayer}
    \end{scope}
    %
    \begin{pgfonlayer}{bg}
      \begin{scope}[fill=colElectrode, fill opacity=.5, draw=colGuard]
        \fill (ftl) -- (rtl) -- (rtr) -- (ftr) -- cycle;
        \draw (ftleA) -- (rtleA) -- (rtreA) -- (ftreA) -- cycle;
        \draw (ftleB) -- (rtleB) -- (rtreB) -- (ftreB) -- cycle;
      \end{scope}
      \begin{scope}[draw=colWire, scale=2, thick,]
        \draw
          ($(ftr)!.5!(rtr)+(-5pt,0)$)
          ++(0,.14pt,-1pt) coordinate (corner) -- ++(0.0,0,0)
          coordinate [at end] (neg) -- ++(0, -.1, 0)
          coordinate (pos) -- ++(0,-.1,0) -- ++ (.3, .15, 0)
          coordinate (tip) -- ++ (-.3, .15, 0) -- ++ (0, -.1, 0);
        \draw (pos) ++(1.3pt,.8pt) -- ($(pos)+(1.3pt,-.8pt)$);
        \draw (pos) ++(.5pt,0pt) -- ($(pos)+(2.1pt,0pt)$);
        \draw (neg) ++(.5pt,0pt) -- ($(neg)+(2.1pt,0pt)$);
        \draw (pos) -| ++(-4pt,-4pt) edge +(-1.5pt, 0) -- ++(1.5pt,0);
        \draw[{Circle[length=2pt,width=2pt]}-{Circle[length=2pt,width=2pt]}]%
          ([yshift=-.4pt, xshift=-2pt]corner) -- ++(0,5.8pt) coordinate (tial)
          -| coordinate (tiar) ([xshift=2pt,yshift=-.4pt]tip);
        \draw[-] (tip) -- ++(5pt, 0) node [draw, right] (li) {Lock-In};
        \draw[-{Circle[open]}] (li.south) -- ([yshift=-5pt]li.south)
          node [at end, below] {$V_{\text{out}}$};
        \draw (neg) -| ($(ftl)!.5!(rtr)$);
        \node [minimum width=15pt,minimum height=8pt, draw=colWire, fill=white,
            inner sep=0cm, outer sep=0cm] (tiares) at ($(tial)!.5!(tiar)$) {};
        \node [above,yshift=2pt] at (tiares) {$R_{\text{TIA}}= \SI{1}{\giga\ohm}$};
      \end{scope}
    \end{pgfonlayer}
    \begin{pgfonlayer}{bg2}
      \begin{scope}[draw=colWire,thick]
        \draw [-{Circle[open]}]
          ($(fbl)!.5!(rbr)$) -- ++(0,-.15,0) -- ++(0,0,2) -- ++(-.5,0,0)
          node [at end, left] {$V_{\text{offset}}$};
      \end{scope}
      \begin{scope}[colElectrode, opacity=.5]
        \fill (fbl) -- (rbl) -- (rbr) -- (fbr) -- cycle;
      \end{scope}
    \end{pgfonlayer}
    %
    \begin{scope}[thick]
      \coordinate (base) at ($(-\cellx/2-1, .62)+(5pt,0)$);
      \draw (base) -- ++(12pt, 0)
        node [at start, left] {$\gs$};
      \draw ([yshift=10pt]base) -- ++(12pt, 0)
        node [at start, left] {$\as$};
      \draw ([yshift=20pt]base) -- ++(12pt, 0)
        node [at start, left] {$\hs$};
      \draw ([yshift=30pt]base) -- ++(12pt, 0)
        node [at start, left] {$\ryd$};
      \foreach [count=\c] \i/\col/\lbl in
      {%
        0/colLaserUV/\SI{226}{\nm},%
        1/colLaserGreen/\SI{540}{\nm},%
        2/colLaserRed/835 -- \SI{855}{\nm}%
      }{%
        \draw [-{Stealth},draw=\col, shorten <=.5mm, shorten >=.5mm] %
          ($(base)+(6pt,\i*10pt)$) --
          ($(base)+(6pt,\c*10pt)$) node [midway, right] {\lbl};
      }%
    \end{scope}
  \end{scope}
  %
  \begin{pgfonlayer}{bg}
    \path (gasstart) ++(-25pt,0) coordinate (gstartbend) ++(0,-15pt)
      coordinate (gstart);
    \path (gasend) ++(25pt,0) coordinate (gendbend) ++(0,25pt)
      coordinate (gend);
    \begin{scope}[line width=4pt,draw=colGas,opacity=.5,transparency group]
      \draw[-{Stealth[flex, length=.5cm]}]%
        (gstart) .. controls (gstartbend) .. (gasstart)
        -- (gasend) .. controls (gendbend) .. (gend);%
    \end{scope}
    \node [below, xshift=17pt, yshift=3pt,align=left] at (gstart)
      {gas mixtures};
  \end{pgfonlayer}
  %
\end{tikzpicture}

%% file: content/results.tex
\begin{figure}[b]
  \includegraphics{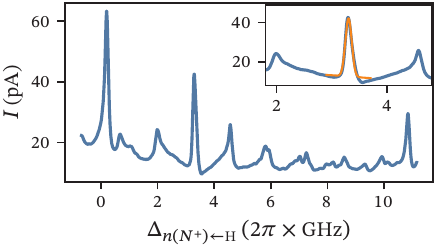}
  \caption{%
    Exemplary measurement trace of a pure NO measurement of 32(4) at a density of $N_{\text{NO}}\approx \SI{1.91+-0.57e15}{\per\cm\cubed}$.
    The inset shows the fit of the non--moving centered line, which is used to obtain the relevant data.
    The fit aligns well with the overall shape of the peak, and is as such sufficient to extract both the FWHM and its frequency position.
    \label{fig:ExampleTrace}%
  }
\end{figure}
\section{Results}
For the evaluation of our measurements we start by calculating the frequency axis out of the time axis by using the known free--spectral range of the Rydberg laser's transfer cavity.
The peak distances are fitted against the position in time by using a third--order polynomial function.
The voltage signal is smoothed by a filter such that a following data reduction is not affecting the quality.
Note that the \tia~converts and amplifies from current to voltage, and the signal we process here is the lock--in amplifiers output signal.
For each trace the most dominant center peak of the high--$l$--manifold, i.e.~the peak not affected by the Stark effect, and the left and
right of it are fitted using individual Voigt functions of the form
\begin{equation}
  f(\omega) = A\cdot f_\text{Voigt} \left(\omega,\omega_0,\gamma_\text{L},\sigma\right)
  +\mathcal{P}_3(\omega) \quad ,
\end{equation}
where $\mathcal{P}_3(\omega)$ is a polynomial function of third order to account for the baseline, $A$ is the amplitude, $\gamma_\text{L}$ is the Lorentzian part of the full width at half maximum (\fwhm), $\gamma_\text{G} = 2\sqrt{2\log(2)}\sigma$ is the Gaussian part of the \fwhm, and $\omega_{0}$ is the center position relevant to extract the shift.
The Voigt function $f_\text{Voigt}$ is normalized to its amplitude.
Special care is taken for the polynomial bounds such that the effects on the parameters of interest are negligible.
While we only look at the center peak, the side peaks are fitted for consistency checking.
In \FigRef{fig:ExampleTrace} we show an exemplary measurement trace including a fit.
\begin{figure*}[htb]
  \includegraphics{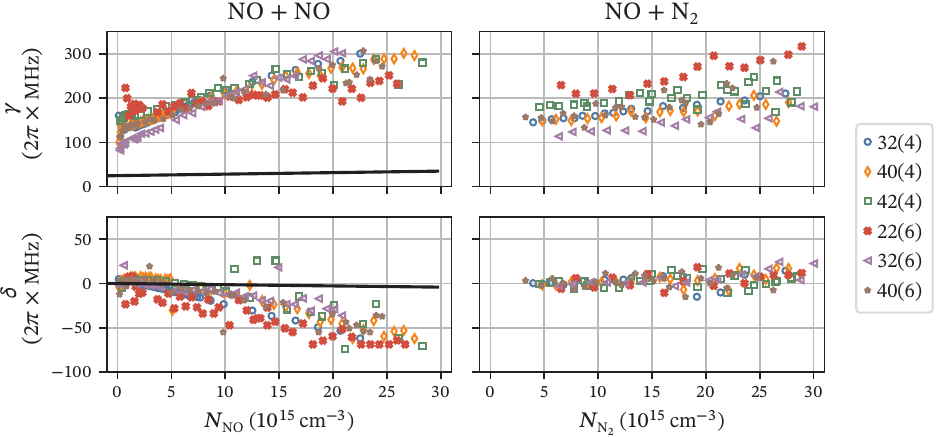}
  \caption{%
    \fwhm~$\gamma$ and relative shift $\delta$ for \no~being subject to an increasing density \density~of the background gas, either \no~itself on the left or \nitrogen~on the right.
    Plotted are the results for different principal quantum numbers $n$ and rotational quantum numbers $N^{+}$ of the ionic core (Hund's case (d)), and the notation $n(N^{+})$ is used in the legend.
    The relative shift is moved such that the first measurement points are referenced to zero.
    The solid line indicates possible contributions by effects on the \hs--state as explained in the text.
    For the measurements of NO perturbed by \nitrogen~the NO density was \SI{0.6+-0.3e15}{\per\cm\cubed}.
    \label{fig:matrix}%
  }
\end{figure*}

We use
\begin{equation}
  \gamma \approx \frac{1}{2} \left[1.0692 \frac{\gamma_\text{L}}{2}+\sqrt{0.86639\left(\frac{\gamma_\text{L}}{2}\right)^2+4\gamma_\text{G}^2}\right]
\end{equation}
from \cite{Olivero1977} to calculate the overall \fwhm~$\gamma$.
Note that we would expect a Lorentzian shape due to the homogeneity of collisional broadening.
However, as we will see later, additional contributions such as the small pressure shift of the second transition \tranGreen~are present in our experiment.
During the evaluation we verified that a fit with a Voigt function suits our results best.
In \FigRef{fig:matrix} we show our results for the \fwhm~$\gamma$ and the relative shift $\delta$.

We start by considering the pressure broadening of the selected Rydberg state $n(N^{+})$.
In general, several mechanisms contribute to the broadening of the linewidth $\gamma$ of the Rydberg line, namely residual Doppler broadening, power broadening, transit-time broadening and broadening due to polarization effects.
In the case of \no~it is hard to give numbers to Doppler broadening, power broadening or polarization effects as literature values were inconsistent \cite{Piper1986,Callear1963,deVivie1988}.
However, in the case of Doppler broadening, a lower bound can be given by taking the linewidth of $\angfreq{15}{\mega\hertz}$ as seen in \cite{Kaspar2022} of the \uv~transition and accounting for the $k$--vector mismatch, which yields around \angfreq{8}{\mega\hertz}.
Additionally, the linear behavior shown in \FigRef{fig:matrix} suggests that collisional effects are dominating.
In the case of collisions of \no~with \no~or \no~with \nitrogen~the broadening is basically the same with increasing density of the perturber.
\nitrogen~and \no~differ only by a single electron when considering the orbital structure.
For \no~only one of the two levels of the $\pi^{*}_{2\text{p}}$ is occupied with a single electron \cite{schulzweilingPhD}, whereas for \nitrogen~all are empty.
This makes the broadening contributions similar.
On top the Boltzmann distribution, describing the population of the rotational levels for a diatomic molecule, is given by Herzberg as \cite{HerzbergMol1}
\begin{equation}
  \label{eq:rotPopulation}
  P(J) = hc B_v \frac{2J+1}{k_{\text{B}}T}\exp
  \left[ -hc
    \frac{B_{v}J(J+1)}{k_\text{B}T}
    \right] \quad ,
\end{equation}
where $P(J)$ is the population probability of having a rotational state with total angular momentum $J$ at temperature $T$ occupied.
The Planck constant is denoted by $h$, $k_{\text{B}}$ is the Boltzmann constant and $c$ is the speed of light.
For \no~the rotational constant $B_v$, as given in \cite{danielak1997} yields, that rotational levels from $J=0.5$ to $J=19.5$ have a population probability above $\SI{1}{\percent}$, suggesting significant contribution to the broadening.
At a later point we will compare our results to measurements done in alkalis.
It is worth to note, that any alkali being a simple atom, lacks these additional contributions \cite{Fuechtbauer1934,WeberNiemax1982}.
Consider the relative shift $\delta$ next.
Based on Fermi's work \cite{Fermi1934} elastic collisions between the Rydberg electron and perturbing atoms or molecules lead to a frequency shift $\delta$ of
\begin{equation}
  \delta = 2\pi\frac{\hbar}{m_{e}} a \cdot \density,\label{eq:shiftFrequency}
\end{equation}
where $\density$ denotes the density, $m_{e}$ is the electronic mass, $a$ is the scattering length, and $\hbar$ is the reduced Planck constant.
In the case of \no~perturbed by \no~a red shift can be seen.
In the case of perturbations by \nitrogen~the considered density range indicates a slight shift to the blue less than the scattering of the measured values for the different states $n(N^+)$.
Inelastic collisions do not lead to a phase shift in the Rydberg electron's wavefunction but change the state of the Rydberg atom or molecule itself.
While both collision types may occur, this gives an indication in likeliness.
For a rare gas such elastic collisions are expected due to the closed--shell structure \cite{Fuechtbauer1934,WeberNiemax1982}.
Interestingly our results show similar results for perturbations of \no~with \no~itself.
A possible explanation for this is that \no~has a degenerate level in the $\pi^{*}_{2\text{p}}$ available, suggesting that elastic collisions may occur despite the rotational and vibrational freedom.
This explanation is supported by the fact that the molecule \no$^{-}$, though short-lived, even exists in biological processes \cite{Hughes1999}.
However, when looking at \nitrogen, a similar argument could be made.
From our measurements it is undecidable, if a particular collision type is dominating for both cases.
Our experiment is only sensitive to the effective scattering length, which can be the average of positive and negative scattering lengths for each channel.

We performed similar measurements without the Rydberg laser to investigate the effect on the \hs--state.
This means we locked the lowest transition, scanned the \tranGreen{} transition and read out electronically.
Note, that the applied field of about \SI{11.9}{\volt\per\cm} is not enough to produce a visible Stark splitting.
Hence, the evaluation focused on a single peak, but was done in the same way as for the Rydberg transition otherwise.
While a broadening and shift could be observed, its contribution is negligible.
We show this in an added solid line to \FigRef{fig:matrix}, where we scaled a linear fit to these measurements by the wavevector mismatch as was done in equation (4) in \cite{Kuebler2018}:
\begin{equation}
  \frac{k_1 - k_2 - k_3}{k_1 - k_2} \approx \num{0.53}
\end{equation}

The resulting scaled functions are %
\begin{subequations}
  \begin{align}
    \gamma (\density) / 2\pi & \approx \SI{0.34e-15}{\mega\hertz\cm\cubed} \cdot \density + \SI{24.87}{\mega\hertz}, \\
    \delta(\density) / 2\pi  & \approx \SI{-0.14e-15}{\mega\hertz\cm\cubed}\cdot \density .
  \end{align}
\end{subequations}
While we do not have to consider this contribution for the overall behavior, the contribution by the intermediate transition \tranGreen~partly explains the necessity of using a Voigt function for fitting rather than a Lorentzian.
Further inhomogeneities might arise from the already introduced available degrees of freedom in a molecule.
\begin{table}[t]
  \caption{%
    Selected examplary values at different principal quantum numbers $n$ of Rydberg states in alkalis from measurements taken by Weber et al.~\cite{WeberNiemax1982} and Füchtbauer et al.~\cite{Fuechtbauer1934}.
    For measurements by Füchtbauer, Rydberg states of potassium (K) are perturbed by argon (Ar), and for measurements by Weber we take their values of rubidium (Rb) Rydberg $n$S--states perturbed by argon.
    In the case of Füchtbauer we calculated the rates by using the ideal gas law.
    \label{tab:alkaliValues}
  }
  \begin{tabular}
    {%
    r||%
    >{\raggedleft\arraybackslash}p{3cm}||%
    >{\raggedleft\arraybackslash}p{1.5cm}|>{\raggedleft\arraybackslash}p{1.5cm}%
    }
        & \multicolumn{1}{c||}{Füchtbauer et al.}                                 & \multicolumn{2}{c}{Weber et al.}                                                          \\
        & \multicolumn{1}{c||}{K + Ar}                                            & \multicolumn{2}{c}{($n$S)\,Rb + Ar}                                                       \\
    $n$ & $\delta/\density$                                                       & $\gamma/\density$                                                     & $\delta/\density$ \\
        & \multicolumn{1}{c||}{$(10^{-15}\,\,2\pi\times\text{MHz}\,\text{cm}^3)$} & \multicolumn{2}{c}{$(10^{-15}\,\,2\pi\times\text{MHz}\,\text{cm}^3)$}                     \\
    \hline
    21  & -10.97                                                                  & 1.46                                                                  & -4.76             \\
    23  & -10.94                                                                  & 1.56                                                                  & -4.94             \\
    25  & -11.34                                                                  & 1.66                                                                  & -5.02             \\
    27  &                                                                         & 1.77                                                                  & -5                \\
    29  &                                                                         & 1.8                                                                   & -5                \\
    33  &                                                                         & 1.36                                                                  & -5.16             \\
    35  &                                                                         & 1.36                                                                  & -5.04             \\
  \end{tabular}
\end{table}
\begin{table}[ht]
  \caption{%
    Broadening and shift rate, $\gamma/\density$ and $\delta/\density$, extracted from a fit to the measurement results shown in \FigRef{fig:matrix}.
    The fit functions are explained in the text.
    The main error contribution to the values is the absolute pressure uncertainty of \SI{30}{\percent} of the used gauges.
    A plot of these values is shown in \FigRef{fig:fuechtbauerNONO}.
    \label{tab:combined}
  }
  \begin{tabular}
    {%
    r||%
    >{\raggedleft\arraybackslash}p{1.5cm}|>{\raggedleft\arraybackslash}p{1.5cm}||%
    >{\raggedleft\arraybackslash}p{1.5cm}|>{\raggedleft\arraybackslash}p{1.5cm}%
    }
    State      & \multicolumn{2}{c||}{NO + NO}                                           & \multicolumn{2}{c}{NO + N\textsubscript{2}}                                                                    \\
    $n(N^{+})$ & $\gamma/\density$                                                       & $\delta/\density$                                                     & $\gamma/\density$ & $\delta/\density$  \\
               & \multicolumn{2}{c||}{$(10^{-15}\,\,2\pi\times\text{MHz}\,\text{cm}^3)$} & \multicolumn{2}{c}{$(10^{-15}\,\,2\pi\times\text{MHz}\,\text{cm}^3)$}                                          \\
    \hline
    $32(4)$    & \num{13(4)}                                                             & \num{-2.8(0.8)}                                                       & \num{5.5(1.6)}    & \num{0.25(0.08)}   \\
    $40(4)$    & \num{11(3)}                                                             & \num{-2.5(0.7)}                                                       & \num{4.0(1.2)}    & \num{0.57(0.17)}   \\
    $42(4)$    & \num{9.1(2.7)}                                                          & \num{-2.1(0.6)}                                                       & \num{4.0(1.2)}    & \num{0.19(0.06)}   \\
    $22(6)$    & \num{7.5(2.3)}                                                          & \num{-2.7(0.8)}                                                       & \num{9.2(2.8)}    & \num{0.52(0.16)}   \\
    $32(6)$    & \num{15(4)}                                                             & \num{-1.4(0.4)}                                                       & \num{4.9(1.5)}    & \num{0.82(0.25)}   \\
    $40(6)$    & \num{10(3)}                                                             & \num{-2.3(0.7)}                                                       & \num{6.1(1.8)}    & \num{0.089(0.027)} \\
  \end{tabular}
\end{table}

In a next step we compare our results with results from the literature.
Füchtbauer et al.~\cite{Fuechtbauer1934} and Weber et al.~\cite{WeberNiemax1982} both investigated the shift of spectroscopic lines of alkalis being subject to a perturbing rare gas.
In Füchtbauer's experiment the alkalis sodium and potassium were subject to collisions with three different rare gases, helium, neon and argon.
In Weber's experiment the rubidium $n$D and $n$S transitions were shifted by helium, xenon and argon.
In contrast to Füchtbauer, Weber was able to analyze the broadening as well.
Comparison is easiest when using the broadening rate $\gamma/\density$ and shift rate $\delta/\density$ as introduced by Weber.
In \TblRef{tab:alkaliValues} we list some exemplary literature values.
We give both values extracted from a fit to the data plotted in \FigRef{fig:matrix} and listed in \TblRef{tab:combined}.
The fit of the shift against the density is done by a linear function in accordance with \EqnRef{eq:shiftFrequency}.
For the broadening we use
\begin{equation}
  f_{\text{fit}} (\density) = \sqrt{\text{offset}^{2} + \left(\frac{\gamma}{\density} \cdot \density\right)^{2}} \quad ,
\end{equation}
as a fit function to account for additional broadening contributions such as the residual Doppler broadening in the low densities by an offset.
We give an error estimate to all fitted values as well.
The largest contribution to our errors has its origin in the absolute pressure given by our pressure gauges (Pfeiffer PKR~251) and amounts to \SI{30}{\percent}.
In contrast, as seen in \cite{Kaspar2022}, the uncertainty of our frequency axis, $\angfreq{2.5}{\mega\hertz}$, can in this case be neglected.
As such the error given is based solely on the pressure uncertainty.
In comparison with \cite{WeberNiemax1982,Fuechtbauer1934} our broadening rate $\gamma/\density$ is several times higher than their values.
The shift rate $\delta/\density$ deviates at least by a factor of two, when comparing measurements of \no~and alkalis, which indicates that additional processes are present in molecules.
\begin{figure}[ht]
  \includegraphics{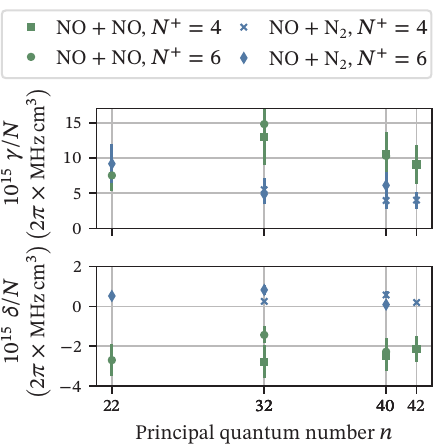}%
  \caption{%
    Broadening rate $\gamma/\density$ and shift rate $\delta/\density$ of the perturbation of Rydberg states in \no.
    We explain the high broadening rate in comparison to alkalis as seen in \cite{Fuechtbauer1934,WeberNiemax1982} by the additional degrees of freedom in a molecule.
    Some exemplary alkali values are given in \ref{tab:alkaliValues}.
    An uncertainty of \SI{30}{\percent} in the measured pressure is indicated by error bars.
    For some measurement points the marker covers the error bar.
    Within the error the values are constant, which is expected when compared to alkalis.
    The individual values are given in \TblRef{tab:combined}.
    \label{fig:fuechtbauerNONO}
  }
\end{figure}

%% file: content/conclusion.tex
\section{Summary}
We showed the collisional shift and broadening of Rydberg states in \no~with increasing density of two background gases.
The detection of the Rydberg states was realized by an electrical readout of the current generated by free charges resulting from collisions.
The linear shift of elastic collisions can be understood as introduced by Fermi \cite{Fermi1934}.

The measurement was either performed on \no~being perturbed by itself or \nitrogen.
Comparison to literature values \cite{Fuechtbauer1934,WeberNiemax1982} of the experiment was achieved by extracting the broadening rate $\gamma/\density$ and shift rate $\delta/\density$ using fits to our experimental data.
The analysis showed that we are unable to determine in our experiment which collision type is dominating.
In any case our rates deviate at least by a factor of two, when compared to measurements done in alkalis.
We attribute this to the additional degrees of freedom in \no.

The overall project's goal is to realize a breath-gas sensor for \no~in the context of a possible medical application.
Our work shows that the main effect to consider is the broadening rate, since it is larger than the shift rate.
This allows for an optimal choice of background densities depending on the excitation bandwidths given by the lasers and other broadening mechanisms present in the system, e.g.~power broadening.
Another significant step towards the realization is the reduction of the cross--sensitivity to other gases present in the breath.
Here, we could use our current setup, but change the background gas for example to CO\textsubscript{2}.
%
\section*{Acknowledgements}
This project has received funding from the European Union's Horizon 2020 research and innovation program under Grant Agreement No.~820393 (macQsimal) as well as by the Deutsche Forschungsgemeinschaft (DFG, German Research Foundation) 431314977/GRK2642 (Research Training Group:
``Towards Graduate Experts in Photonic Quantum Technologies'').
Additionally, we would like to thank Prof.~S.~Hogan, M.~Rayment, Prof.~H.~Sadeghpour and Prof.~R.~González Férez for fruitful discussions and valuable advice.

%% file: main.bbl
\begin{thebibliography}{31}%
\makeatletter
\providecommand \@ifxundefined [1]{%
 \@ifx{#1\undefined}
}%
\providecommand \@ifnum [1]{%
 \ifnum #1\expandafter \@firstoftwo
 \else \expandafter \@secondoftwo
 \fi
}%
\providecommand \@ifx [1]{%
 \ifx #1\expandafter \@firstoftwo
 \else \expandafter \@secondoftwo
 \fi
}%
\providecommand \natexlab [1]{#1}%
\providecommand \enquote  [1]{``#1''}%
\providecommand \bibnamefont  [1]{#1}%
\providecommand \bibfnamefont [1]{#1}%
\providecommand \citenamefont [1]{#1}%
\providecommand \href@noop [0]{\@secondoftwo}%
\providecommand \href [0]{\begingroup \@sanitize@url \@href}%
\providecommand \@href[1]{\@@startlink{#1}\@@href}%
\providecommand \@@href[1]{\endgroup#1\@@endlink}%
\providecommand \@sanitize@url [0]{\catcode `\\12\catcode `\$12\catcode
  `\&12\catcode `\#12\catcode `\^12\catcode `\_12\catcode `\%12\relax}%
\providecommand \@@startlink[1]{}%
\providecommand \@@endlink[0]{}%
\providecommand \url  [0]{\begingroup\@sanitize@url \@url }%
\providecommand \@url [1]{\endgroup\@href {#1}{\urlprefix }}%
\providecommand \urlprefix  [0]{URL }%
\providecommand \Eprint [0]{\href }%
\providecommand \doibase [0]{https://doi.org/}%
\providecommand \selectlanguage [0]{\@gobble}%
\providecommand \bibinfo  [0]{\@secondoftwo}%
\providecommand \bibfield  [0]{\@secondoftwo}%
\providecommand \translation [1]{[#1]}%
\providecommand \BibitemOpen [0]{}%
\providecommand \bibitemStop [0]{}%
\providecommand \bibitemNoStop [0]{.\EOS\space}%
\providecommand \EOS [0]{\spacefactor3000\relax}%
\providecommand \BibitemShut  [1]{\csname bibitem#1\endcsname}%
\let\auto@bib@innerbib\@empty
\bibitem [{\citenamefont {Arnold}\ \emph {et~al.}(1977)\citenamefont {Arnold},
  \citenamefont {Mittal}, \citenamefont {Katsuki},\ and\ \citenamefont
  {Murad}}]{Arnold1977}%
  \BibitemOpen
  \bibfield  {author} {\bibinfo {author} {\bibfnamefont {W.~P.}\ \bibnamefont
  {Arnold}}, \bibinfo {author} {\bibfnamefont {C.~K.}\ \bibnamefont {Mittal}},
  \bibinfo {author} {\bibfnamefont {S.}~\bibnamefont {Katsuki}},\ and\ \bibinfo
  {author} {\bibfnamefont {F.}~\bibnamefont {Murad}},\ }\bibfield  {title}
  {\bibinfo {title} {Nitric oxide activates guanylate cyclase and increases
  guanosine 3 \textquotesingle:5\textquotesingle-cyclic monophosphate levels in
  various tissue preparations},\ }\href
  {https://doi.org/10.1073/pnas.74.8.3203} {\bibfield  {journal} {\bibinfo
  {journal} {Proceedings of the National Academy of Sciences}\ }\textbf
  {\bibinfo {volume} {74}},\ \bibinfo {pages} {3203} (\bibinfo {year}
  {1977})}\BibitemShut {NoStop}%
\bibitem [{\citenamefont {Furchgott}\ and\ \citenamefont
  {Zawadzki}(1980)}]{Furchgott1980}%
  \BibitemOpen
  \bibfield  {author} {\bibinfo {author} {\bibfnamefont {R.~F.}\ \bibnamefont
  {Furchgott}}\ and\ \bibinfo {author} {\bibfnamefont {J.~V.}\ \bibnamefont
  {Zawadzki}},\ }\bibfield  {title} {\bibinfo {title} {The obligatory role of
  endothelial cells in the relaxation of arterial smooth muscle by
  acetylcholine},\ }\href {https://doi.org/10.1038/288373a0} {\bibfield
  {journal} {\bibinfo  {journal} {Nature}\ }\textbf {\bibinfo {volume} {288}},\
  \bibinfo {pages} {373} (\bibinfo {year} {1980})}\BibitemShut {NoStop}%
\bibitem [{\citenamefont {Ignarro}\ \emph {et~al.}(1987)\citenamefont
  {Ignarro}, \citenamefont {Buga}, \citenamefont {Wood}, \citenamefont
  {Byrns},\ and\ \citenamefont {Chaudhuri}}]{Ignarro1987}%
  \BibitemOpen
  \bibfield  {author} {\bibinfo {author} {\bibfnamefont {L.~J.}\ \bibnamefont
  {Ignarro}}, \bibinfo {author} {\bibfnamefont {G.~M.}\ \bibnamefont {Buga}},
  \bibinfo {author} {\bibfnamefont {K.~S.}\ \bibnamefont {Wood}}, \bibinfo
  {author} {\bibfnamefont {R.~E.}\ \bibnamefont {Byrns}},\ and\ \bibinfo
  {author} {\bibfnamefont {G.}~\bibnamefont {Chaudhuri}},\ }\bibfield  {title}
  {\bibinfo {title} {Endothelium-derived relaxing factor produced and released
  from artery and vein is nitric oxide.},\ }\href
  {https://doi.org/10.1073/pnas.84.24.9265} {\bibfield  {journal} {\bibinfo
  {journal} {Proceedings of the National Academy of Sciences}\ }\textbf
  {\bibinfo {volume} {84}},\ \bibinfo {pages} {9265} (\bibinfo {year}
  {1987})}\BibitemShut {NoStop}%
\bibitem [{\citenamefont {{The Nobel Foundation}}(1998)}]{NobelPrize}%
  \BibitemOpen
  \bibfield  {author} {\bibinfo {author} {\bibnamefont {{The Nobel
  Foundation}}},\ }\href
  {https://www.nobelprize.org/prizes/medicine/1998/summary/} {\bibinfo {title}
  {The nobel prize in physiology or medicine}} (\bibinfo {year}
  {1998})\BibitemShut {NoStop}%
\bibitem [{\citenamefont {Ignarro}(2018)}]{Ignarro2018}%
  \BibitemOpen
  \bibfield  {author} {\bibinfo {author} {\bibfnamefont {L.~J.}\ \bibnamefont
  {Ignarro}},\ }\bibfield  {title} {\bibinfo {title} {Nitric oxide is not just
  blowing in the wind},\ }\href {https://doi.org/10.1111/bph.14540} {\bibfield
  {journal} {\bibinfo  {journal} {British Journal of Pharmacology}\ }\textbf
  {\bibinfo {volume} {176}},\ \bibinfo {pages} {131} (\bibinfo {year}
  {2018})}\BibitemShut {NoStop}%
\bibitem [{\citenamefont {Haklar}\ \emph {et~al.}(2001)\citenamefont {Haklar},
  \citenamefont {Sayin-\"{O}zveri}, \citenamefont {Y\"{u}ksel}, \citenamefont
  {Aktan},\ and\ \citenamefont {Yal{\c{c}}in}}]{Haklar2001}%
  \BibitemOpen
  \bibfield  {author} {\bibinfo {author} {\bibfnamefont {G.}~\bibnamefont
  {Haklar}}, \bibinfo {author} {\bibfnamefont {E.}~\bibnamefont
  {Sayin-\"{O}zveri}}, \bibinfo {author} {\bibfnamefont {M.}~\bibnamefont
  {Y\"{u}ksel}}, \bibinfo {author} {\bibfnamefont {A.}~\bibnamefont {Aktan}},\
  and\ \bibinfo {author} {\bibfnamefont {A.}~\bibnamefont {Yal{\c{c}}in}},\
  }\bibfield  {title} {\bibinfo {title} {Different kinds of reactive oxygen and
  nitrogen species were detected in colon and breast tumors},\ }\href
  {https://doi.org/10.1016/s0304-3835(01)00421-9} {\bibfield  {journal}
  {\bibinfo  {journal} {Cancer Letters}\ }\textbf {\bibinfo {volume} {165}},\
  \bibinfo {pages} {219} (\bibinfo {year} {2001})}\BibitemShut {NoStop}%
\bibitem [{\citenamefont {(Choudhari)}\ \emph {et~al.}(2012)\citenamefont
  {(Choudhari)}, \citenamefont {Sridharan}, \citenamefont {Gadbail},\ and\
  \citenamefont {Poornima}}]{KordeChoudhari2012}%
  \BibitemOpen
  \bibfield  {author} {\bibinfo {author} {\bibfnamefont {S.~K.}\ \bibnamefont
  {(Choudhari)}}, \bibinfo {author} {\bibfnamefont {G.}~\bibnamefont
  {Sridharan}}, \bibinfo {author} {\bibfnamefont {A.}~\bibnamefont {Gadbail}},\
  and\ \bibinfo {author} {\bibfnamefont {V.}~\bibnamefont {Poornima}},\
  }\bibfield  {title} {\bibinfo {title} {Nitric oxide and oral cancer: A
  review},\ }\href {https://doi.org/10.1016/j.oraloncology.2012.01.003}
  {\bibfield  {journal} {\bibinfo  {journal} {Oral Oncology}\ }\textbf
  {\bibinfo {volume} {48}},\ \bibinfo {pages} {475} (\bibinfo {year}
  {2012})}\BibitemShut {NoStop}%
\bibitem [{\citenamefont {Xu}\ \emph {et~al.}(2002)\citenamefont {Xu},
  \citenamefont {Liu}, \citenamefont {Loizidou}, \citenamefont {Ahmed},\ and\
  \citenamefont {Charles}}]{XU2002}%
  \BibitemOpen
  \bibfield  {author} {\bibinfo {author} {\bibfnamefont {W.}~\bibnamefont
  {Xu}}, \bibinfo {author} {\bibfnamefont {L.~Z.}\ \bibnamefont {Liu}},
  \bibinfo {author} {\bibfnamefont {M.}~\bibnamefont {Loizidou}}, \bibinfo
  {author} {\bibfnamefont {M.}~\bibnamefont {Ahmed}},\ and\ \bibinfo {author}
  {\bibfnamefont {I.~G.}\ \bibnamefont {Charles}},\ }\bibfield  {title}
  {\bibinfo {title} {The role of nitric oxide in cancer},\ }\href
  {https://doi.org/10.1038/sj.cr.7290133} {\bibfield  {journal} {\bibinfo
  {journal} {Cell Research}\ }\textbf {\bibinfo {volume} {12}},\ \bibinfo
  {pages} {311} (\bibinfo {year} {2002})}\BibitemShut {NoStop}%
\bibitem [{\citenamefont {Khan}\ \emph {et~al.}(2020)\citenamefont {Khan},
  \citenamefont {Dervan}, \citenamefont {Bhattacharyya}, \citenamefont
  {McAuliffe}, \citenamefont {Miranda},\ and\ \citenamefont
  {Glynn}}]{Khan2020}%
  \BibitemOpen
  \bibfield  {author} {\bibinfo {author} {\bibfnamefont {F.~H.}\ \bibnamefont
  {Khan}}, \bibinfo {author} {\bibfnamefont {E.}~\bibnamefont {Dervan}},
  \bibinfo {author} {\bibfnamefont {D.~D.}\ \bibnamefont {Bhattacharyya}},
  \bibinfo {author} {\bibfnamefont {J.~D.}\ \bibnamefont {McAuliffe}}, \bibinfo
  {author} {\bibfnamefont {K.~M.}\ \bibnamefont {Miranda}},\ and\ \bibinfo
  {author} {\bibfnamefont {S.~A.}\ \bibnamefont {Glynn}},\ }\bibfield  {title}
  {\bibinfo {title} {The role of nitric oxide in cancer: Master regulator or
  {NOt}?},\ }\href {https://doi.org/10.3390/ijms21249393} {\bibfield  {journal}
  {\bibinfo  {journal} {International Journal of Molecular Sciences}\ }\textbf
  {\bibinfo {volume} {21}},\ \bibinfo {pages} {9393} (\bibinfo {year}
  {2020})}\BibitemShut {NoStop}%
\bibitem [{\citenamefont {Thomas}\ \emph {et~al.}(2008)\citenamefont {Thomas},
  \citenamefont {Ridnour}, \citenamefont {Isenberg}, \citenamefont
  {Flores-Santana}, \citenamefont {Switzer}, \citenamefont {Donzelli},
  \citenamefont {Hussain}, \citenamefont {Vecoli}, \citenamefont {Paolocci},
  \citenamefont {Ambs}, \citenamefont {Colton}, \citenamefont {Harris},
  \citenamefont {Roberts},\ and\ \citenamefont {Wink}}]{Thomas2008}%
  \BibitemOpen
  \bibfield  {author} {\bibinfo {author} {\bibfnamefont {D.~D.}\ \bibnamefont
  {Thomas}}, \bibinfo {author} {\bibfnamefont {L.~A.}\ \bibnamefont {Ridnour}},
  \bibinfo {author} {\bibfnamefont {J.~S.}\ \bibnamefont {Isenberg}}, \bibinfo
  {author} {\bibfnamefont {W.}~\bibnamefont {Flores-Santana}}, \bibinfo
  {author} {\bibfnamefont {C.~H.}\ \bibnamefont {Switzer}}, \bibinfo {author}
  {\bibfnamefont {S.}~\bibnamefont {Donzelli}}, \bibinfo {author}
  {\bibfnamefont {P.}~\bibnamefont {Hussain}}, \bibinfo {author} {\bibfnamefont
  {C.}~\bibnamefont {Vecoli}}, \bibinfo {author} {\bibfnamefont
  {N.}~\bibnamefont {Paolocci}}, \bibinfo {author} {\bibfnamefont
  {S.}~\bibnamefont {Ambs}}, \bibinfo {author} {\bibfnamefont {C.~A.}\
  \bibnamefont {Colton}}, \bibinfo {author} {\bibfnamefont {C.~C.}\
  \bibnamefont {Harris}}, \bibinfo {author} {\bibfnamefont {D.~D.}\
  \bibnamefont {Roberts}},\ and\ \bibinfo {author} {\bibfnamefont {D.~A.}\
  \bibnamefont {Wink}},\ }\bibfield  {title} {\bibinfo {title} {The chemical
  biology of nitric oxide: Implications in cellular signaling},\ }\href
  {https://doi.org/10.1016/j.freeradbiomed.2008.03.020} {\bibfield  {journal}
  {\bibinfo  {journal} {Free Radical Biology and Medicine}\ }\textbf {\bibinfo
  {volume} {45}},\ \bibinfo {pages} {18} (\bibinfo {year} {2008})}\BibitemShut
  {NoStop}%
\bibitem [{\citenamefont {Gustafsson}\ \emph {et~al.}(1991)\citenamefont
  {Gustafsson}, \citenamefont {Leone}, \citenamefont {Persson}, \citenamefont
  {Wiklund},\ and\ \citenamefont {Moncada}}]{Gustafsson1991}%
  \BibitemOpen
  \bibfield  {author} {\bibinfo {author} {\bibfnamefont {L.}~\bibnamefont
  {Gustafsson}}, \bibinfo {author} {\bibfnamefont {A.}~\bibnamefont {Leone}},
  \bibinfo {author} {\bibfnamefont {M.}~\bibnamefont {Persson}}, \bibinfo
  {author} {\bibfnamefont {N.}~\bibnamefont {Wiklund}},\ and\ \bibinfo {author}
  {\bibfnamefont {S.}~\bibnamefont {Moncada}},\ }\bibfield  {title} {\bibinfo
  {title} {Endogenous nitric oxide is present in the exhaled air of rabbits,
  guinea pigs and humans},\ }\href
  {https://doi.org/10.1016/0006-291x(91)91268-h} {\bibfield  {journal}
  {\bibinfo  {journal} {Biochemical and Biophysical Research Communications}\
  }\textbf {\bibinfo {volume} {181}},\ \bibinfo {pages} {852} (\bibinfo {year}
  {1991})}\BibitemShut {NoStop}%
\bibitem [{\citenamefont {{American Thoracic Society}}\ and\ \citenamefont
  {{European Respiratory Society}}(2005)}]{ATS2005a}%
  \BibitemOpen
  \bibfield  {author} {\bibinfo {author} {\bibnamefont {{American Thoracic
  Society}}}\ and\ \bibinfo {author} {\bibnamefont {{European Respiratory
  Society}}},\ }\bibfield  {title} {\bibinfo {title} {{ATS}/{ERS}
  recommendations for standardized procedures for the online and offline
  measurement of exhaled lower respiratory nitric oxide and nasal nitric oxide,
  2005},\ }\href {https://doi.org/10.1164/rccm.200406-710st} {\bibfield
  {journal} {\bibinfo  {journal} {American Journal of Respiratory and Critical
  Care Medicine}\ }\textbf {\bibinfo {volume} {171}},\ \bibinfo {pages} {912}
  (\bibinfo {year} {2005})}\BibitemShut {NoStop}%
\bibitem [{\citenamefont {Schmidt}\ \emph {et~al.}(2018)\citenamefont
  {Schmidt}, \citenamefont {Fiedler}, \citenamefont {Albrecht}, \citenamefont
  {Djekic}, \citenamefont {Schalberger}, \citenamefont {Baur}, \citenamefont
  {Löw}, \citenamefont {Fruehauf}, \citenamefont {Pfau}, \citenamefont
  {Anders}, \citenamefont {Grant},\ and\ \citenamefont
  {Kübler}}]{Schmidt2018}%
  \BibitemOpen
  \bibfield  {author} {\bibinfo {author} {\bibfnamefont {J.}~\bibnamefont
  {Schmidt}}, \bibinfo {author} {\bibfnamefont {M.}~\bibnamefont {Fiedler}},
  \bibinfo {author} {\bibfnamefont {R.}~\bibnamefont {Albrecht}}, \bibinfo
  {author} {\bibfnamefont {D.}~\bibnamefont {Djekic}}, \bibinfo {author}
  {\bibfnamefont {P.}~\bibnamefont {Schalberger}}, \bibinfo {author}
  {\bibfnamefont {H.}~\bibnamefont {Baur}}, \bibinfo {author} {\bibfnamefont
  {R.}~\bibnamefont {Löw}}, \bibinfo {author} {\bibfnamefont {N.}~\bibnamefont
  {Fruehauf}}, \bibinfo {author} {\bibfnamefont {T.}~\bibnamefont {Pfau}},
  \bibinfo {author} {\bibfnamefont {J.}~\bibnamefont {Anders}}, \bibinfo
  {author} {\bibfnamefont {E.~R.}\ \bibnamefont {Grant}},\ and\ \bibinfo
  {author} {\bibfnamefont {H.}~\bibnamefont {Kübler}},\ }\bibfield  {title}
  {\bibinfo {title} {Proof of concept for an optogalvanic gas sensor for {NO}
  based on rydberg excitations},\ }\bibfield  {journal} {\bibinfo  {journal}
  {Applied Physics Letters}\ }\textbf {\bibinfo {volume} {113}},\ \href
  {https://doi.org/10.1063/1.5024321} {10.1063/1.5024321} (\bibinfo {year}
  {2018})\BibitemShut {NoStop}%
\bibitem [{\citenamefont {Fermi}(1934)}]{Fermi1934}%
  \BibitemOpen
  \bibfield  {author} {\bibinfo {author} {\bibfnamefont {E.}~\bibnamefont
  {Fermi}},\ }\bibfield  {title} {\bibinfo {title} {Sopra lo spostamento per
  pressione delle righe elevate delle serie spettrali},\ }\href
  {https://doi.org/10.1007/bf02959829} {\bibfield  {journal} {\bibinfo
  {journal} {Il Nuovo Cimento}\ }\textbf {\bibinfo {volume} {11}},\ \bibinfo
  {pages} {157} (\bibinfo {year} {1934})}\BibitemShut {NoStop}%
\bibitem [{\citenamefont {Füchtbauer}\ \emph {et~al.}(1934)\citenamefont
  {Füchtbauer}, \citenamefont {Schulz},\ and\ \citenamefont
  {Brandt}}]{Fuechtbauer1934}%
  \BibitemOpen
  \bibfield  {author} {\bibinfo {author} {\bibfnamefont {C.}~\bibnamefont
  {Füchtbauer}}, \bibinfo {author} {\bibfnamefont {P.}~\bibnamefont
  {Schulz}},\ and\ \bibinfo {author} {\bibfnamefont {A.~F.}\ \bibnamefont
  {Brandt}},\ }\bibfield  {title} {\bibinfo {title} {Verschiebung von hohen
  {S}erienlinien des {N}atriums und {K}aliums durch {F}remdgase, {B}erechnung
  der {W}irkungsquerschnitte von {E} delgasen gegen sehr langsame
  {E}lektronen},\ }\href {https://doi.org/10.1007/bf01334059} {\bibfield
  {journal} {\bibinfo  {journal} {Zeitschrift für Physik}\ }\textbf {\bibinfo
  {volume} {90}},\ \bibinfo {pages} {403} (\bibinfo {year} {1934})}\BibitemShut
  {NoStop}%
\bibitem [{\citenamefont {Weber}\ and\ \citenamefont
  {Niemax}(1982)}]{WeberNiemax1982}%
  \BibitemOpen
  \bibfield  {author} {\bibinfo {author} {\bibfnamefont {K.~H.}\ \bibnamefont
  {Weber}}\ and\ \bibinfo {author} {\bibfnamefont {K.}~\bibnamefont {Niemax}},\
  }\bibfield  {title} {\bibinfo {title} {Impact broadening and shift of {Rb nS}
  {and nD} levels by noble gases},\ }\href {https://doi.org/10.1007/bf01416067}
  {\bibfield  {journal} {\bibinfo  {journal} {Zeitschrift für Physik A Atoms
  and Nuclei}\ }\textbf {\bibinfo {volume} {307}},\ \bibinfo {pages} {13}
  (\bibinfo {year} {1982})}\BibitemShut {NoStop}%
\bibitem [{\citenamefont {Gonz{\'{a}}lez-F{\'{e}}rez}\ \emph
  {et~al.}(2021)\citenamefont {Gonz{\'{a}}lez-F{\'{e}}rez}, \citenamefont
  {Shertzer},\ and\ \citenamefont {Sadeghpour}}]{GonzlezFrez2021}%
  \BibitemOpen
  \bibfield  {author} {\bibinfo {author} {\bibfnamefont {R.}~\bibnamefont
  {Gonz{\'{a}}lez-F{\'{e}}rez}}, \bibinfo {author} {\bibfnamefont
  {J.}~\bibnamefont {Shertzer}},\ and\ \bibinfo {author} {\bibfnamefont
  {H.}~\bibnamefont {Sadeghpour}},\ }\bibfield  {title} {\bibinfo {title}
  {Ultralong-range rydberg bimolecules},\ }\bibfield  {journal} {\bibinfo
  {journal} {Physical Review Letters}\ }\textbf {\bibinfo {volume} {126}},\
  \href {https://doi.org/10.1103/physrevlett.126.043401}
  {10.1103/physrevlett.126.043401} (\bibinfo {year} {2021})\BibitemShut
  {NoStop}%
\bibitem [{\citenamefont {Kaspar}\ \emph {et~al.}(2022)\citenamefont {Kaspar},
  \citenamefont {Munkes}, \citenamefont {Neufeld}, \citenamefont {Ebel},
  \citenamefont {Schellander}, \citenamefont {Löw}, \citenamefont {Pfau},\
  and\ \citenamefont {Kübler}}]{Kaspar2022}%
  \BibitemOpen
  \bibfield  {author} {\bibinfo {author} {\bibfnamefont {P.}~\bibnamefont
  {Kaspar}}, \bibinfo {author} {\bibfnamefont {F.}~\bibnamefont {Munkes}},
  \bibinfo {author} {\bibfnamefont {P.}~\bibnamefont {Neufeld}}, \bibinfo
  {author} {\bibfnamefont {L.}~\bibnamefont {Ebel}}, \bibinfo {author}
  {\bibfnamefont {Y.}~\bibnamefont {Schellander}}, \bibinfo {author}
  {\bibfnamefont {R.}~\bibnamefont {Löw}}, \bibinfo {author} {\bibfnamefont
  {T.}~\bibnamefont {Pfau}},\ and\ \bibinfo {author} {\bibfnamefont
  {H.}~\bibnamefont {Kübler}},\ }\href
  {https://doi.org/10.1103/PhysRevA.106.062816} {\bibfield  {journal} {\bibinfo
   {journal} {Phys. Rev. A}\ }\textbf {\bibinfo {volume} {106}},\ \bibinfo
  {pages} {062816} (\bibinfo {year} {2022})}\BibitemShut {NoStop}%
\bibitem [{\citenamefont {Neuhaus}\ \emph {et~al.}(2017)\citenamefont
  {Neuhaus}, \citenamefont {Metzdorff}, \citenamefont {Chua}, \citenamefont
  {Jacqmin}, \citenamefont {Briant}, \citenamefont {Heidmann}, \citenamefont
  {Cohadon},\ and\ \citenamefont {Deléglise}}]{NeuhausPYRPL2017}%
  \BibitemOpen
  \bibfield  {author} {\bibinfo {author} {\bibfnamefont {L.}~\bibnamefont
  {Neuhaus}}, \bibinfo {author} {\bibfnamefont {R.}~\bibnamefont {Metzdorff}},
  \bibinfo {author} {\bibfnamefont {S.}~\bibnamefont {Chua}}, \bibinfo {author}
  {\bibfnamefont {T.}~\bibnamefont {Jacqmin}}, \bibinfo {author} {\bibfnamefont
  {T.}~\bibnamefont {Briant}}, \bibinfo {author} {\bibfnamefont
  {A.}~\bibnamefont {Heidmann}}, \bibinfo {author} {\bibfnamefont {P.-F.}\
  \bibnamefont {Cohadon}},\ and\ \bibinfo {author} {\bibfnamefont
  {S.}~\bibnamefont {Deléglise}},\ }\bibfield  {title} {\bibinfo {title}
  {Pyrpl (python red pitaya lockbox) — an open-source software package for
  fpga-controlled quantum optics experiments},\ }in\ \href
  {https://doi.org/10.1109/CLEOE-EQEC.2017.8087380} {\emph {\bibinfo
  {booktitle} {2017 Conference on Lasers and Electro-Optics Europe \& European
  Quantum Electronics Conference (CLEO/Europe-EQEC)}}}\ (\bibinfo {year}
  {2017})\ pp.\ \bibinfo {pages} {1--1}\BibitemShut {NoStop}%
\bibitem [{\citenamefont {Mäusezahl}\ \emph {et~al.}(tion)\citenamefont
  {Mäusezahl}, \citenamefont {Munkes},\ and\ \citenamefont
  {Löw}}]{laserLockTut}%
  \BibitemOpen
  \bibfield  {author} {\bibinfo {author} {\bibfnamefont {M.}~\bibnamefont
  {Mäusezahl}}, \bibinfo {author} {\bibfnamefont {F.}~\bibnamefont {Munkes}},\
  and\ \bibinfo {author} {\bibfnamefont {R.}~\bibnamefont {Löw}},\ }\href@noop
  {} {\bibinfo {title} {Tutorial on locking techniques and the manufacturing of
  vapor cells for spectroscopy}} (\bibinfo {year} {in preparation})\BibitemShut
  {NoStop}%
\bibitem [{\citenamefont {Western}(2017)}]{Western2017}%
  \BibitemOpen
  \bibfield  {author} {\bibinfo {author} {\bibfnamefont {C.~M.}\ \bibnamefont
  {Western}},\ }\bibfield  {title} {\bibinfo {title} {{PGOPHER}: A program for
  simulating rotational, vibrational and electronic spectra},\ }\href
  {https://doi.org/10.1016/j.jqsrt.2016.04.010} {\bibfield  {journal} {\bibinfo
   {journal} {Journal of Quantitative Spectroscopy and Radiative Transfer}\
  }\textbf {\bibinfo {volume} {186}},\ \bibinfo {pages} {221} (\bibinfo {year}
  {2017})}\BibitemShut {NoStop}%
\bibitem [{\citenamefont {Danielak}\ \emph {et~al.}(1997)\citenamefont
  {Danielak}, \citenamefont {Domin}, \citenamefont {Ke}, \citenamefont
  {Rytel},\ and\ \citenamefont {Zachwieja}}]{danielak1997}%
  \BibitemOpen
  \bibfield  {author} {\bibinfo {author} {\bibfnamefont {J.}~\bibnamefont
  {Danielak}}, \bibinfo {author} {\bibfnamefont {U.}~\bibnamefont {Domin}},
  \bibinfo {author} {\bibfnamefont {R.}~\bibnamefont {Ke}}, \bibinfo {author}
  {\bibfnamefont {M.}~\bibnamefont {Rytel}},\ and\ \bibinfo {author}
  {\bibfnamefont {M.}~\bibnamefont {Zachwieja}},\ }\bibfield  {title} {\bibinfo
  {title} {Reinvestigation of the emission $\gamma$ band system ($\as$ -- $
  \gsgeneric$) of the no molecule},\ }\href
  {https://doi.org/https://doi.org/10.1006/jmsp.1996.7181} {\bibfield
  {journal} {\bibinfo  {journal} {Journal of Molecular Spectroscopy}\ }\textbf
  {\bibinfo {volume} {181}},\ \bibinfo {pages} {394} (\bibinfo {year}
  {1997})}\BibitemShut {NoStop}%
\bibitem [{\citenamefont {Ogi}\ \emph {et~al.}(2000)\citenamefont {Ogi},
  \citenamefont {Takahashi}, \citenamefont {Tsukiyama},\ and\ \citenamefont
  {Bersohn}}]{Ogi2000}%
  \BibitemOpen
  \bibfield  {author} {\bibinfo {author} {\bibfnamefont {Y.}~\bibnamefont
  {Ogi}}, \bibinfo {author} {\bibfnamefont {M.}~\bibnamefont {Takahashi}},
  \bibinfo {author} {\bibfnamefont {K.}~\bibnamefont {Tsukiyama}},\ and\
  \bibinfo {author} {\bibfnamefont {R.}~\bibnamefont {Bersohn}},\ }\bibfield
  {title} {\bibinfo {title} {Laser-induced amplified spontaneous emission from
  the 3d and nf rydberg states of {NO}},\ }\href
  {https://doi.org/10.1016/s0301-0104(00)00043-4} {\bibfield  {journal}
  {\bibinfo  {journal} {Chemical Physics}\ }\textbf {\bibinfo {volume} {255}},\
  \bibinfo {pages} {379} (\bibinfo {year} {2000})}\BibitemShut {NoStop}%
\bibitem [{\citenamefont {Olivero}\ and\ \citenamefont
  {Longbothum}(1977)}]{Olivero1977}%
  \BibitemOpen
  \bibfield  {author} {\bibinfo {author} {\bibfnamefont {J.}~\bibnamefont
  {Olivero}}\ and\ \bibinfo {author} {\bibfnamefont {R.}~\bibnamefont
  {Longbothum}},\ }\bibfield  {title} {\bibinfo {title} {Empirical fits to the
  voigt line width: A brief review},\ }\href
  {https://doi.org/10.1016/0022-4073(77)90161-3} {\bibfield  {journal}
  {\bibinfo  {journal} {Journal of Quantitative Spectroscopy and Radiative
  Transfer}\ }\textbf {\bibinfo {volume} {17}},\ \bibinfo {pages} {233}
  (\bibinfo {year} {1977})}\BibitemShut {NoStop}%
\bibitem [{\citenamefont {Piper}\ and\ \citenamefont
  {Cowles}(1986)}]{Piper1986}%
  \BibitemOpen
  \bibfield  {author} {\bibinfo {author} {\bibfnamefont {L.~G.}\ \bibnamefont
  {Piper}}\ and\ \bibinfo {author} {\bibfnamefont {L.~M.}\ \bibnamefont
  {Cowles}},\ }\bibfield  {title} {\bibinfo {title} {Einstein coefficients and
  transition moment variation for the {NO (A \,$ {}^2\Sigma^+$--X\,${}^2\Pi$)}
  transition},\ }\href {https://doi.org/10.1063/1.451098} {\bibfield  {journal}
  {\bibinfo  {journal} {The Journal of Chemical Physics}\ }\textbf {\bibinfo
  {volume} {85}},\ \bibinfo {pages} {2419–2422} (\bibinfo {year}
  {1986})}\BibitemShut {NoStop}%
\bibitem [{\citenamefont {Callear}\ and\ \citenamefont
  {Smith}(1963)}]{Callear1963}%
  \BibitemOpen
  \bibfield  {author} {\bibinfo {author} {\bibfnamefont {A.~B.}\ \bibnamefont
  {Callear}}\ and\ \bibinfo {author} {\bibfnamefont {I.~W.~M.}\ \bibnamefont
  {Smith}},\ }\bibfield  {title} {\bibinfo {title} {Fluorescence of nitric
  oxide. part 1.—determination of the mean lifetime of the {A\,$
  {}^2\Sigma^+$} state},\ }\href {https://doi.org/10.1039/tf9635901720}
  {\bibfield  {journal} {\bibinfo  {journal} {Trans. Faraday Soc.}\ }\textbf
  {\bibinfo {volume} {59}},\ \bibinfo {pages} {1720–1734} (\bibinfo {year}
  {1963})}\BibitemShut {NoStop}%
\bibitem [{\citenamefont {de~Vivie}\ and\ \citenamefont
  {Peyerimhoff}(1988)}]{deVivie1988}%
  \BibitemOpen
  \bibfield  {author} {\bibinfo {author} {\bibfnamefont {R.}~\bibnamefont
  {de~Vivie}}\ and\ \bibinfo {author} {\bibfnamefont {S.~D.}\ \bibnamefont
  {Peyerimhoff}},\ }\bibfield  {title} {\bibinfo {title} {Theoretical
  spectroscopy of the no radical. i. potential curves and lifetimes of excited
  states},\ }\href {https://doi.org/10.1063/1.454958} {\bibfield  {journal}
  {\bibinfo  {journal} {The Journal of Chemical Physics}\ }\textbf {\bibinfo
  {volume} {89}},\ \bibinfo {pages} {3028–3043} (\bibinfo {year}
  {1988})}\BibitemShut {NoStop}%
\bibitem [{\citenamefont {Schulz-Weiling}(2017)}]{schulzweilingPhD}%
  \BibitemOpen
  \bibfield  {author} {\bibinfo {author} {\bibfnamefont {M.}~\bibnamefont
  {Schulz-Weiling}},\ }\emph {\bibinfo {title} {Ultracold Molecular Plasma}},\
  \href@noop {} {Ph.D. thesis} (\bibinfo {year} {2017})\BibitemShut {NoStop}%
\bibitem [{\citenamefont {Herzberg}(1950)}]{HerzbergMol1}%
  \BibitemOpen
  \bibfield  {author} {\bibinfo {author} {\bibfnamefont {G.}~\bibnamefont
  {Herzberg}},\ }\href@noop {} {\emph {\bibinfo {title} {Spectra of Diatomic
  Molecules}}},\ \bibinfo {edition} {second edition}\ ed.,\ \bibinfo {series}
  {Molecular Spectra and Molecular Structure}, Vol.~\bibinfo {volume} {1}\
  (\bibinfo  {publisher} {D. Van Nostrand Company, Inc.},\ \bibinfo {year}
  {1950})\BibitemShut {NoStop}%
\bibitem [{\citenamefont {Hughes}(1999)}]{Hughes1999}%
  \BibitemOpen
  \bibfield  {author} {\bibinfo {author} {\bibfnamefont {M.~N.}\ \bibnamefont
  {Hughes}},\ }\bibfield  {title} {\bibinfo {title} {Relationships between
  nitric oxide, nitroxyl ion, nitrosonium cation and peroxynitrite},\ }\href
  {https://doi.org/10.1016/s0005-2728(99)00019-5} {\bibfield  {journal}
  {\bibinfo  {journal} {Biochimica et Biophysica Acta ({BBA}) - Bioenergetics}\
  }\textbf {\bibinfo {volume} {1411}},\ \bibinfo {pages} {263} (\bibinfo {year}
  {1999})}\BibitemShut {NoStop}%
\bibitem [{\citenamefont {K\"{u}bler}\ and\ \citenamefont
  {Shaffer}(2018)}]{Kuebler2018}%
  \BibitemOpen
  \bibfield  {author} {\bibinfo {author} {\bibfnamefont {H.}~\bibnamefont
  {K\"{u}bler}}\ and\ \bibinfo {author} {\bibfnamefont {J.~P.}\ \bibnamefont
  {Shaffer}},\ }\bibfield  {title} {\bibinfo {title} {A read-out enhancement
  for microwave electric field sensing with rydberg atoms},\ }in\ \href
  {https://doi.org/10.1117/12.2309386} {\emph {\bibinfo {booktitle} {Quantum
  Technologies 2018}}},\ \bibinfo {editor} {edited by\ \bibinfo {editor}
  {\bibfnamefont {A.~J.}\ \bibnamefont {Shields}}, \bibinfo {editor}
  {\bibfnamefont {J.}~\bibnamefont {Stuhler}},\ and\ \bibinfo {editor}
  {\bibfnamefont {M.~J.}\ \bibnamefont {Padgett}}}\ (\bibinfo  {publisher}
  {{SPIE}},\ \bibinfo {year} {2018})\BibitemShut {NoStop}%
\end{thebibliography}%
